**Work-Life Balance Starts with Proper Deadlines and Exemplary Agencies**

Noé Lugaz[1], Réka M. Winslow[1], Nada Al-Haddad[1], Christina O. Lee[2], Sarah K. Vines[3], Katharine Reeves[4], Amir Caspi[5], Daniel Seaton[5], Cooper Downs[6], Lindsay Glesener[7], Angelos Vourlidas[3], Camilla Scolini[1], Tibor Török[6], Robert Allen[3], Erika Palmerio[6]

[1]Space Science Center, University of New Hampshire; [2]Space Science Laboratory, University of California – Berkeley, [3]Johns Hopkins University/Applied Physics Laboratory, [4]Center for Astrophysics, Harvard Smithsonian, [5]Southwest Research Institute, [6]Predictive Science Inc., [7]University of Minnesota

**Synopsis**: Diversity, equity and inclusion (DEI) programs can only be implemented successfully if proper work-life balance is possible in Heliophysics (and in STEM field in general). One of the core issues stems from the culture of "work-above-life" associated with mission concepts, development, and implementation but also the expectations that seem to originate from numerous announcements from NASA (and other agencies). The benefits of work-life balance are well documented; however, the entire system surrounding research in Heliophysics hinders or discourages proper work-life balance. For example, there does not seem to be attention paid by NASA Headquarters (HQ) on the timing of their announcements regarding how it will be perceived by researchers, and how the timing may promote a culture where work trumps personal life. The same is true for remarks by NASA HQ program officers during panels or informal discussions, where seemingly innocuous comments may give a perception that work is expected after "normal" work hours. In addition, we are calling for work-life balance plans and implementation to be one of the criteria used for down-selection and confirmation of missions (Key Decision Points: KDP-B, KDP-C).

The benefits of proper work-life balance are well documented in peer-reviewed publications for more than 30 years (e.g., see Crosby, 1991). They include benefits to the individual, their family but also to the employers themselves, in terms of increased productivity, worker retention, improved recruitment capabilities and increased loyalty, among others (see Sirgy & Lee, 2017). This white paper does not dive into this large body of research but invite the Decadal Survey committee to hear from researchers from the social sciences if they need further arguments. Instead, we detail some of the current organization of research in Heliophysics that is detrimental to work-life balance and offer potential solutions.

At the core of work-life balance is the peer pressure and perceived (or real) expectations that individual researchers feel in their work in Heliophysics. Numerous small, seemingly inconsequential announcements and decisions can create an overall culture of "work first, life second." Agencies should carefully consider the ramifications of their announcements, even informal, before releasing them and train program officers for careful considerations of any announcement related to short deadline or late/long work hours, even in "informal settings" (for example while running a panel). In addition, the move to partially online/remote-work has made many of existing issues significantly more prominent and created new issues with work-life balance. We discuss a few examples below. We invite agencies to consider the consequences of deadlines and announcements. We propose using the actual implementation of work-life balance plans during a mission Phases A and B as one of the selection criteria for missions during key decision points (KDPs). This would ensure that the culture of "a successful mission takes a 24/7 work commitment" doesn't continue into the next decades.

**Example 1: Overall Work Culture**
Many informal discussions between the lead author and graduate students and post-doctoral researchers at COSPAR 2022 centered around questions such as "can you be successful while taking at least one day off every week?" (where week is the 7-day week) or "do you have nightmares about your work?". Discussions within the group of authors of this white paper highlighted that these are common questions and extend to later career stages. Overall, this highlights that issues of work culture and work-life balance within STEM in general and geospace/heliophysics science in particular, are especially dire. This issue is compounded by multiple developments: a) The increase over the past decade of data science jobs and engineering jobs in "New Space" companies with high pay; b) The decrease in effective (inflation-adjusted) salary for many early-career scientists, associated with salary increases from universities or laboratories that fall below the level of inflation and increases in cost-of-living (especially rent, which has increased beyond inflation). Taken together, early-career scientists are often confronted with the prospect of having spent around a decade in higher education to have a low-paid job with non-existent work-life balance.
*Solutions:*
- Agencies should use their weights with universities and laboratories to ensure that proper work-life balance strategies are put in place. This should start where agencies have most control (NSF and NPP post-doctoral fellows, NASA employees), extend to contractors for NASA centers. However, efforts should be made that researchers primarily supported by grants and contracts from NASA, NSF and other agencies are able to maintain a proper work-life balance.

- A work-life balance plan should be required for large proposals (instrument development, mission, strategic capabilities).
- As it is unlikely that university salaries will become competitive with those of employees in data science or new space companies, the only way to keep the "best and brightest" within science is to ensure that they at least have a better work-life balance than in these other jobs. The current solution of "only those who are truly motivated by the love of science will stay" is both inadequate and inequitable, especially if we are to increase the number of first-generation students and researchers from underserved minorities within Heliophysics.
- Agencies should fund state-of-the-profession surveys, taking advantage of NSF ongoing research in education and social sciences. Some professional societies (e.g., AGU, AAS) already do this, but this information needs to flow up to the funding agencies directly. This should be used to set up specific goals for work-life balance which could be reviewed during decadal survey and mid-term reviews.
- There should be specific guidelines as part of grants and contracts for researchers funded primarily through grants and contracts (post-doctoral and "soft-money" researchers). This could include requirements that holidays and vacations are paid through indirect costs (rather than charged as direct cost to the grant, often well beyond the date at which those costs have accrued). Agencies could also recommend (or require, if legally possible) that employers offer explicit time allocations for professional activities outside of grants, funded by indirect cost, to provide funded time for proposal-writing and other career development tasks that are readily available to "hard-money" researchers but not usually to those on "soft" money.
- Whenever possible (see below), agencies should lead by example.

**Example 2: Work Culture Surrounding Instruments and Missions**
Participation in a mission is often seen as the ultimate achievement in Heliophysics. Early- and mid-career researchers are often welcomed within instrument or mission teams with "badge of honor/horror" stories, such as "that day we ran our first EM testing during Christmas," or "remember this week of overnighters we pulled before the proposal submission." Such stories are obviously detrimental to the perception of work requirement and work-life balance in STEM, and what it takes to run a successful mission or build a successful instrument. For early- and mid-career researchers proposing or participating in their first mission, at the Mission of Opportunity or Explorer level, the expectations in terms of proposal preparations are often daunting unless they are being mentored actively by researchers who have had leadership role within missions, or unless they are parts of centers (such as NASA centers or APL) where there is significant level of internal support to develop mission concepts (like the presence of the APL Concurrent Engineering, or ACE, laboratory). Mentoring is often offered through "patronage," where researchers that are being mentored to become future instrument leads or part of mission management are chosen because they are at the right institute at the right time. While such a model might be appropriate for instrument leads, where there is an institutional heritage, it is not appropriate for other mission roles, such as PI, deputy PI and Project/Mission Scientist, and science working group lead, who do not require institutional heritage.

*Solutions:*

- Review work-life balance as part of Key Decision Points (KDPs) for missions. At a minimum, this would be a review of the work-life balance plan submitted within the mission proposal and its implementation. It may require changes in the scope of the Standard Review Board (SRB) Handbook and the addition of members in the SRB that can assess this plan and its implementation. This could also be in the form of additional reviews if there is staff turnover over a given threshold, to see if this is simply due to staff leaving for more exciting projects or if this is related to burnout or a caustic work culture in the team. Having a high-level requirement within a mission regarding work-life balance that can be tracked (employee turnover, number of vacation days taken) is another venue that could be taken.
- Ensure that holidays and vacation time is integrated into the instrument and mission master schedule and that key reviews are not held during or just after major holidays (Thanksgiving, end-of-the year). This could be connected to the work-life balance plan. Heliophysics missions often do not have extremely tight launch constraints, as is the case for Planetary missions (for which, targeting a specific planet, moon or planetary body requires a launch in a tight window). For example, it is possible to launch to L1 more than 3 weeks every month. As such, Heliophysics mission schedule can more easily take into consideration work-life balance. NASA should be willing to pay for the additional costs, if any, associated with a slightly longer schedule.
- Integrate not only a DEI Plan but also a training plan for instrument leads, PIs, deputies, and project managers regarding work-life balance as part of the proposal.
- Develop ways for a lower "cost" of entry (in term of work-life balance) into missions for early-career scientists, maybe through the development of SIMPLEX-like concept. While reducing the number of pages for the proposal is tempting, it can have the opposite effect of requiring a more polished proposal unless the risk tolerance of TMC maturity is further decreased for step-1 selections. While the "cost" in terms of proposal preparation could be similar for a proposal submitted to SIMPLEX as compared to H-FORT, the higher budget for SIMPLEX could reduce the pressure on the PI and Project Manager (PM) to juggle multiple projects.
- NASA should also consider offering access to laboratories for concept studies (like for the decadal survey concept studies) after a pre-selection based on science and/or following the SIMPLEX example of funding a Phase A/B so that there is a reward to be selected for a concept study (rather than having to invest more "unpaid" time for a Phase A concept study). This latter option also improves equity for proposers from institutions that do not have the internal resources to invest in unpaid mission design exercises, or prior mission heritage – allowing the science to be driven by the "right" institution by partnering with existing mission design houses.

**Example 3: Schedule of Remote Review Panels**

Before COVID, most NASA panels used to be in person in the greater DC area or in San Antonio, TX. Expectations of work in the evening used to make (some) sense when researchers were staying in a hotel away from their family. The same is, however, not true for remote panels where the panelists are logging in from their home. The problem is compounded by the three to six time zones (including Hawaii and Alaska) that the United States spans. Some NASA panels have been starting at 7am PT or running beyond 7pm ET with the expectation that researchers will do additional writing after the "day" is finished. First, this gives a wrong impression from NASA Headquarters program officers regarding the expectations that NASA has in term of work schedule, since panels are one of the only ways where early and mid-career scientists interact with NASA HQ program officers. Second, this puts exceptional pressure on early and mid-career scientists, especially (but not only) women who are more often in charge of childcare and have a larger share of housework. Participation in panels is voluntary and eligible panelists are given a modest honorarium. However, participation in panels is seen by many early-career scientists as a necessary a) to understand the selection process, b) to build up their resume. While remote participation in panels eliminates travel time and provides more equitable access to panel participation by people who cannot easily travel (e.g., due to family obligations), care must be taken to ensure that remote participation is not more onerous than in-person.

*Solution:*
- Adapt the schedule of remote panels to account for the fact that panelists are at their home where they can be expected to have family and housework duties. This includes reducing the expectation of "homework" to be done in the evenings.
- Have more asynchronous work time during panels to limit "in person" work to time periods from about 10:30am to 5:30pm ET. Adding 1 or 2 interstitial days of asynchronous work between synchronous meetings provides flexibility in scheduling without significantly lengthening the total panel time. While this would also require increased honoraria, this is balanced by significantly reduced travel expenses. Honoraria should ensure that they are sufficient to offset either missed work hours or to cover additional dependent-care expenses required to allow panel participation. This will require training of the program officers and coordination with the panel chairs.
- Ensure that Agencies' program officers receive proper training in work-life balance, focusing specifically on the consequences of announcement and perceived expectations.

**Example 4: The Decadal Survey White Paper Deadline and Other Announcements**
The white paper deadline was announced on June 24 with a deadline of August 18 (pushed since then to August 24 and to September 7). The initial time period for white paper preparation was under 8 weeks (and under 7 when considering delays in broad announcements to the community) and corresponds to the time period of a) school summer holidays (which last from June 10 to August 22 in Fairfax county, VA for example), b) heavy conference schedule with COSPAR, TESS/SPD, the IAU General Assembly and the SHINE workshop, among others during the 2-month time period, c) the release of the Heliophysics SMEX draft AO in June 22, and the deadline for the Heliophysics Flight Opportunities and Technology and Instrument Development for Science in August 31/September 1, d) the time when many researchers do take vacation, especially after two years of restricted travel due to the COVID situation.

While this is just one example, schedules for other announcements have also been hard to understand. For example, the community announcement for the 2022 SMEX and Mission of Opportunity call was sent at 2pm ET on 2021 December 23[rd]. As the final AO was originally to be released in June 2022, it is unclear why the initial announcement could not wait until early 2022 but had to instead occur just before a major holiday season. In addition, the final AO initially expected in June 2022 was then changed to August, then changed to September. As of this WP deadline, the SMEX 2022 proposal target date is now planned to be due in December 2022. Forward planning for things like conference attendance (e.g., AGU Fall meeting, which is taking place less than a week after the current deadline), and scientific collaborations for people involved in large mission proposals is impossible when projected deadlines for these proposals are constantly shifting, as has happened with the current round of Heliophysics SMEX call. This also applies to holidays and vacations, for which planning is nearly impossible.

*Solution:*
- In general, announcements should be avoided on Friday afternoon or just before a holiday.
- Significant deadlines (the white paper one, the 2022 SMEX ones) are often rumored to be forthcoming for months before the actual announcement. Agencies and NASEM should make more use of official announcements of a forthcoming deadline within a particular quarter (say summer 2022 for the white paper, fall 2022 for the SMEX). This would enable early-career researchers or researchers without a developed network to 1) be aware of the deadline and 2) for those without dedicated institutional funding, to prepare more adequately for this critical deadline. In general, the agencies and academies should announce a worst-case (rather than a hope-for) deadline so that they can stick to the deadline as much as possible without constant changes. In general, a longer time between the announcement and the deadline might be advisable, as it gives more time for researchers to prepare and set their own schedule.
- We propose that major recurring proposals (Explorers, H-FORT, Strategic capabilities, DRIVE centers, etc.) are given a set of 4-6 yearly and fixed deadlines (for example March 1, May 1, July 1, September 1, November 1) and that if there is a shift in the schedule, it is to the next deadline. As NASA is already able to release the yearly ROSES omnibus on February 14 every year, we are confident that the agencies would be able to stick to these fixed deadlines. All announcements on the schedule should be early. NASA should consider the cost-benefit in increasing the time interval between the final AO publication and due date from 90 days to 120 or beyond.

**References:**

Crosby, F. J. (1991). *Juggling: the unexpected advantages of balancing career and home for women and their families*. New York: Free Press.

Sirgy, M., Lee, DJ. *Work-Life Balance: an Integrative Review*. *Applied Research Quality Life* **13**, 229–254 (2018). https://doi.org/10.1007/s11482-017-9509-8